\documentclass[epj]{webofc}
\usepackage[utf8]{inputenc}
\usepackage[varg]{txfonts}   
\usepackage{booktabs}
\usepackage{xcolor}
\definecolor{darkred}{rgb}{0.4,0.0,0.0}
\definecolor{darkgreen}{rgb}{0.0,0.4,0.0}
\definecolor{darkblue}{rgb}{0.0,0.0,0.4}
\usepackage[bookmarks,linktocpage,colorlinks,
    linkcolor = darkred,
    urlcolor  = darkblue,
    citecolor = darkgreen]{hyperref}
\usepackage{subfigure}
\usepackage{amsmath}
\usepackage{multirow}
\wocname{EPJ Web of Conferences}
\woctitle{Lattice2017}
\newcommand{\be}{\begin{equation}}
\newcommand{\ee}{\end{equation}}
\DeclareMathOperator{\erfc}{erfc}
\usepackage{tikzscale}
\usepackage{pgfplots}
\pgfplotsset{compat=1.12}

\usetikzlibrary{arrows.meta}
\usetikzlibrary{shapes.misc}
\usetikzlibrary{backgrounds}
\usetikzlibrary{positioning}
\usetikzlibrary{matrix}
\usepgfplotslibrary{groupplots}

\usetikzlibrary{external}
\tikzsetexternalprefix{fig/external/}

\newcommand{\matrixopfile}{}

\newcommand{\figdir}{./fig}
\newcommand{\datadir}{./data}
\newcommand{\autocorr}{3} 

\definecolor{msd_1}{HTML}{396AB1}
\definecolor{msd_2}{HTML}{DA7C30}
\definecolor{msd_3}{HTML}{3E9651}
\definecolor{msd_4}{HTML}{CC2529}
\definecolor{msd_5}{HTML}{535154}

\pgfplotscreateplotcyclelist{mstone_d}{%
{msd_1, mark=square},
{msd_2, mark=triangle},
{msd_3, mark=o},
{msd_4, mark=diamond},
{msd_5, mark=pentagon},
{color={rgb,255:red,107:green,76:blue,154}},
{color={rgb,255:red,146:green,36:blue,40}},
{color={rgb,255:red,148:green,139:blue,61}}}

\pgfplotsset{
	y errors/.style={
		error bars/y dir = both,
		error bars/y explicit,
	}
}

\pgfplotsset{
	rotated xticklabels/.style={
		xticklabel style={
			rotate=45,
			anchor=east,
			/pgf/number format/precision=4,
		},
	}
}

\pgfplotsset{
	slanted xlabels/.style={
		xticklabel style={
			rotate=45,
			anchor=north east,
			xticklabel shift=-3pt,
			/pgf/number format/precision=4,
		},
	}
}

\pgfplotsset{
	normal plot/.style={
		color=black,
		cycle list={{red, mark=*}},
	}
}

\pgfplotsset{
	faded plot/.style={
		color=black!30!white,
		cycle list={{red!30!white, mark=*}},
	}
}

\pgfplotsset{
	emph plot/.style={
		color=black,
		cycle list={{red, mark=*}},
		axis background/.style={fill=yellow!30!white}, 
	}
}

\pgfplotsset{
	mark1/.style={
		red,
		mark=*
	}
}

\pgfplotsset{
	mark2/.style={
		blue!70!white,
		mark=*
	}
}
\begin{document}
%
\selectlanguage{english}
\title{Single flavour filtering for RHMC in BQCD}
\author{%
\firstname{Waseem} \lastname{Kamleh}\inst{1}\fnsep\thanks{Speaker, \email{waseem.kamleh@adelaide.edu.au}}
\firstname{Taylor}  \lastname{Haar}\inst{1}\and
\firstname{Yoshifumi} \lastname{Nakamura}\inst{2} \and
\firstname{James M.}  \lastname{Zanotti}\inst{1}\fnsep
}
\institute{%
Centre for the Subatomic Structure of Matter, Department of Physics, University of Adelaide, Australia \and
RIKEN Advanced Institute for Computational Science, Kobe, Hyogo 650-0047, Japan
}
\abstract{%
Filtering algorithms for two degenerate quark flavours have advanced to the point that, in 2+1 flavour simulations, the cost of the strange quark is significant compared with the light quarks. This makes efficient filtering algorithms for single flavour actions highly desirable, in particular when considering 1+1+1 flavour simulations for QED+QCD. Here we discuss methods for filtering the RHMC algorithm that are implemented within BQCD, an open-source Fortran program for Hybrid Monte Carlo simulations.
}
\maketitle
\section{Introduction}\label{intro}

While the lattice QCD community has achieved dynamical simulations at
or near the physical pion mass~\cite{Aoki:2008sm,Durr:2008zz,Durr:2010vn,Durr:2010aw},
the Monte Carlo generation of gauge field configurations with light
quarks presents a numerically intensive challenge, and is only
possible for groups with access to primary tier supercomputing
resources. After several decades, the Hybrid Monte Carlo (HMC)
algorithm~\cite{Duane:1987de} remains the method of choice for
dynamical QCD simulations.

The leading expense in modern dynamical simulations is due to the
fermion determinant. To enable the use of Monte Carlo methods, the
Grassmannian fields which represent the quarks are
integrated out analytically,
\be \prod_f \int D\psi_f D\overline{\psi}_f \, \exp \left[-\overline{\psi}_f M_f \psi_f \right] = \prod_f \det{M_f}, \ee
where $M_f$ is the fermion matrix for the quark flavour $f.$ At this
point the resulting fermion matrix determinant can be estimated
stochastically through the introduction of pseudofermions,
\be \prod_f \det{M_f} = \prod_f \int D\phi_f D\phi_f^*\, \exp \left[- \phi^\dagger_f M_f^{-1} \phi_f \right] \ee
The pseudofermion fields $\phi$ are standard complex numbers and are
therefore computable. The simplest pseudofermion formulation is for
two degenerate quark flavours,
\be (\det{M})^2 = \det(M^\dagger M) = \int D\phi D\phi^*\, \exp \left[- \phi^\dagger (M^\dagger M)^{-1} \phi \right], \ee
where we have taken advantage of the fact that the determinant of the fermion matrix $M$ is real. In
this instance, the determinant is represented by a pseudofermion action
with a positive semi-definite matrix,
\be S_{2f} = \phi^\dagger K^{-1} \phi, \ee 
where $K = M^\dagger M.$ The positive semi-definite property enables
the generation of pseudofermion fields that satisfy the probability
distribution $\rho(\phi) \sim e^{-S_{2f}(\phi)}.$

Single flavour fermion simulations pose additional computational
challenges. For Wilson-type actions, the fermion matrix has a complex
spectrum, and hence cannot be used directly in the pseudofermion
distribution without additional modification. Noting that, if $\det M$ is real and positive we have
\be \det M = \sqrt{\det(M^\dagger M)} = \int D\phi D\phi^*\, \exp \left[- \phi^\dagger (M^\dagger M)^{-\frac{1}{2}} \phi \right]. \ee
The Rational HMC (RHMC) algorithm~\cite{Clark:2006fx,Clark:2006wq} solves this problem by replacing the
fermion matrix $M$ with a rational approximation to the inverse square
root,
\be R(K) \approx K^{-1/2}, \ee
where $K = M^\dag M$ is positive semi-definite, such that the pseudofermion action for a single flavour is
\be S_{1f} = \phi^\dag R(K) \phi. \ee
Typically, $R(K)$ is chosen to be the Zolotarev approximation, which
is the optimal rational polynomial approximation for $R(z) \simeq
z^{-\frac{1}{2}}.$

Due to isospin symmetry being very nearly exact, in pure QCD
simulations the up and down quark are almost always chosen to be
degenerate. This means that in a $2+1$ flavour simulation only the
strange must be treated using RHMC. Historically, the cost of adding
the strange quark to a dynamical simulation with already light up and
down quarks was relatively small. However, there have been a large
number of algorithmic improvements focused on reducing the cost of two
flavour simulations, such that in a modern dynamical simulation the
cost of the including the strange quark is significant.

Moreover, the precision of lattice QCD calculations of certain
quantities has advanced to the point that electromagnetic effects can
no longer be neglected, see e.g.~\cite{Horsley:2015eaa,Horsley:2015vla,Borsanyi:2014jba,Borsanyi:2013lga,Aoki:2012st}. In
these QCD+QED calculations, isospin symmetry is explicitly broken as
the up and down quarks have different charges $q_u \neq q_d.$ This
means that at the physical point, $1+1+1$ flavour simulations are
needed~\cite{Aoki:2012st}, noting that the down and strange quark have
different masses $m_d \neq m_s.$ This motivates us to seek
improvements to the single flavour RHMC algorithm that parallel the
advances made for the degenerate two flavour case.

\section{Filtering algorithms}

The essential principle underlying the Hybrid Monte Carlo algorithm is
the introduction of a fictitious simulation time co-ordinate, along
with the corresponding conjugate momenta and a Hamiltonian describing
the evolution in this new simulation time. The molecular dynamics
evolution approximately preserves the extended Hamiltonian and this
ensures a high acceptance rate for the global Metropolis
accept/reject step that takes place at the end of each trajectory.

There are two sources of computational expense in performing
simulations at light quark mass. Both stem from the fact that the
fermion determinant must be represented stochastically using
pseudofermions with a non-local action that features the inverse of
the fermion matrix. The first is that the condition number of the
fermion matrix increases at light quark mass, causing a large increase
in the number of conjugate gradient iterations required to solve the
linear system.  Solving this system is required at each molecular
dynamics step, which means the numerical expense of the inversions is
the dominant cost for generating dynamical gauge fields. The second is
that the molecular dynamics step-size must be decreased as the quark
mass is reduced in order to maintain control over the rapid
fluctuations driven by the pseudofermion field.

Splitting the pseudofermion action using a filter provides the
opportunity to address both these issues. The application of a filter
$F$ to the fermion action kernel $L$ results in a pseudofermion action with multiple terms,
\be \phi^\dag L \phi \rightarrow \phi_1^\dag F \phi_1 + \phi_2^\dag F^{-1}L \phi_2. \ee
A good choice of filter can have a two fold effect. $F^{-1}$ should
act as a preconditioner for $L,$ such that the cost of evaluating the
molecular dynamics force is reduced. The second benefit comes from the
ability to place the evolution of the two terms on different time
scales, such that the expensive fermion matrix inversions are
performed less often per trajectory.

There are a number of filtering techniques~\cite{Kamleh:2011dc,Hasenbusch:2001ne,Urbach:2005ji,AliKhan:2003br,Luscher:2003vf,Luscher:2005rx}
that have been developed for two flavour simulations, two of which are
most relevant to this discussion. Polynomial
filtering~\cite{Kamleh:2011dc} uses a polynomial approximation to the
inverse as a preconditioner for $L,$ with the advantage that a short
polynomial filter can simultaneously capture the high frequency
dynamics whilst being cheap to evaluate. Utilising appropriately
factored polynomials allows for additional time scales to be
introduced. Mass preconditioning~\cite{Hasenbusch:2001ne,Urbach:2005ji}
uses the lattice Dirac matrix at heavier value of the quark mass
parameter as a filter, and again it is possible to create a tiered
molecular dynamics integration scheme by using successively increasing
values of the hopping parameter $\kappa,$ with simulations at the
physical quark mass using four or more Hasenbuch terms. The
possibility of combining mass preconditioning and polynomial filtering
was recently shown to remove the need to fine tune the Hasenbuch
parameter(s)~\cite{Haar:2016bwe}.

In this work we propose two classes of filter that can be applied to
the Rational Hybrid Monte Carlo action,
\be S_{RHMC} = \phi^\dag R(K) \phi. \ee
Analogously to the two flavour case, we can construct a
polynomial filter for RHMC by choosing a low order polynomial $P(K)
\simeq K^{-1/2}$ that approximates the inverse square root, such that
\be S_{PF-RHMC} = \phi_1^\dag P(K) \phi_1 + \phi_2^\dag P(K)^{-1}R(K) \phi_2. \ee
Given a class of polynomials $P(K)$, such as the Chebyshev
approximations, we only need to tune the polynomial order $p.$ The
polynomial term $P(K)$ is cheap to evaluate. So long as the roots of
$P(K)$ are sufficiently away from zero then we can obtain the filtered
rational polynomial $P(K)^{-1}R(K)$ needed for the remainder term with
minimal additional expense by using a multi-shift solver, and the
filter will reduce the force associated with the remainder term such
that it can be placed on a coarser integration scale.
\begin{table}[t]
  \caption{The coefficients for the Zolotarev rational approximation
    to the inverse square root $R(z^2)$ given by
    equation~(\ref{eq:zolo}), optimised for the range $z \in [5\times 10^{-5},3]$ with rational polynomial order $n=20.$}
  \label{tab:zolo}
  \small
  \centering
  \begin{tabular}{ccc}\toprule
    $k$  & $a_k$ & $b_k$ \\\midrule
    1  & $9.563244545350\times 10^{+01}$ & $2.185152657378\times 10^{+01}$ \\
    2  & $8.388503498802\times 10^{+00}$ & $3.869255332779\times 10^{+00}$ \\
    3  & $1.938402066167\times 10^{+00}$ & $1.011688743145\times 10^{+00}$ \\
    4  & $5.393711779563\times 10^{-01}$ & $2.908314285103\times 10^{-01}$ \\
    5  & $1.577758224234\times 10^{-01}$ & $8.587639032804\times 10^{-02}$ \\
    6  & $4.682604103557\times 10^{-02}$ & $2.555797860752\times 10^{-02}$ \\
    7  & $1.395718122336\times 10^{-02}$ & $7.624224868005\times 10^{-03}$ \\
    8  & $4.165459337102\times 10^{-03}$ & $2.275977093976\times 10^{-03}$ \\
    9  & $1.243636491310\times 10^{-03}$ & $6.795636775098\times 10^{-04}$ \\
    10 & $3.713408528751\times 10^{-04}$ & $2.029168564071\times 10^{-04}$ \\
    11 & $1.108828440206\times 10^{-04}$ & $6.059122761302\times 10^{-05}$ \\
    12 & $3.310947727642\times 10^{-05}$ & $1.809210191033\times 10^{-05}$ \\
    13 & $9.885863174458\times 10^{-06}$ & $5.401564705966\times 10^{-06}$ \\
    14 & $2.951119430158\times 10^{-06}$ & $1.612073222848\times 10^{-06}$ \\
    15 & $8.803512407386\times 10^{-07}$ & $4.805018241491\times 10^{-07}$ \\
    16 & $2.620044683831\times 10^{-07}$ & $1.426073879295\times 10^{-07}$ \\
    17 & $7.736439845055\times 10^{-08}$ & $4.171524002026\times 10^{-08}$ \\
    18 & $2.224004044081\times 10^{-08}$ & $1.160749939012\times 10^{-08}$ \\
    19 & $5.815072459290\times 10^{-09}$ & $2.682241760431\times 10^{-09}$ \\
    20 & $1.029675478286\times 10^{-09}$ & $2.352768004344\times 10^{-10}$ \\\midrule
    \multicolumn{3}{c}{$d_n = 6.411915162135\times 10^{-02}$} \\\bottomrule
  \end{tabular}
\end{table}
The second type of filter we propose is based on the observation that
the class of Zolotarev rational polynomials $R(K) \simeq K^{-1/2}$ can be written down in
an \emph{ordered product} form,
\be  R(K) = d_n \prod_{k=1}^n \frac{(K+a_k)}{(K+b_k),} \label{eq:zolo} \ee
where $a_k, b_k, d_n > 0,$ and the coefficients in the numerator and
denominator strictly decrease as $k$ increases such that $a_k >
a_{k+1}, b_k > b_{k+1}.$ Table~\ref{tab:zolo} provides the
coefficients for the $n=20$ Zolotarev approximation $R(z^2)$ optimised
for the range $z \in [5\times 10^{-5},3]$ as an example.  The ordered product
can then be partitioned into partial factors~\cite{Luscher:2012av,Bruno:2014jqa} defined by
\be R_{i,j}(K) = (d_n)^{\delta_{i,0}} \prod_{k=i+1}^j \frac{(K+a_k)}{(K+b_k)}, \ee
where the overall coefficient $d_n$ is only present if $i=0.$ The
value of this partitioning is revealed when we observe that
truncations of the ordered product $R_{0,t}(K)$ at order $t < n$ also
provide an approximation to the inverse square root,
\be R_{0,t}(K) = d_n\prod_{k=1}^t \frac{(K+a_k)}{(K+b_k)} \simeq K^{-1/2}. \ee
\begin{figure}[tb] 
  \centering
  \includegraphics[width=0.7\textwidth]{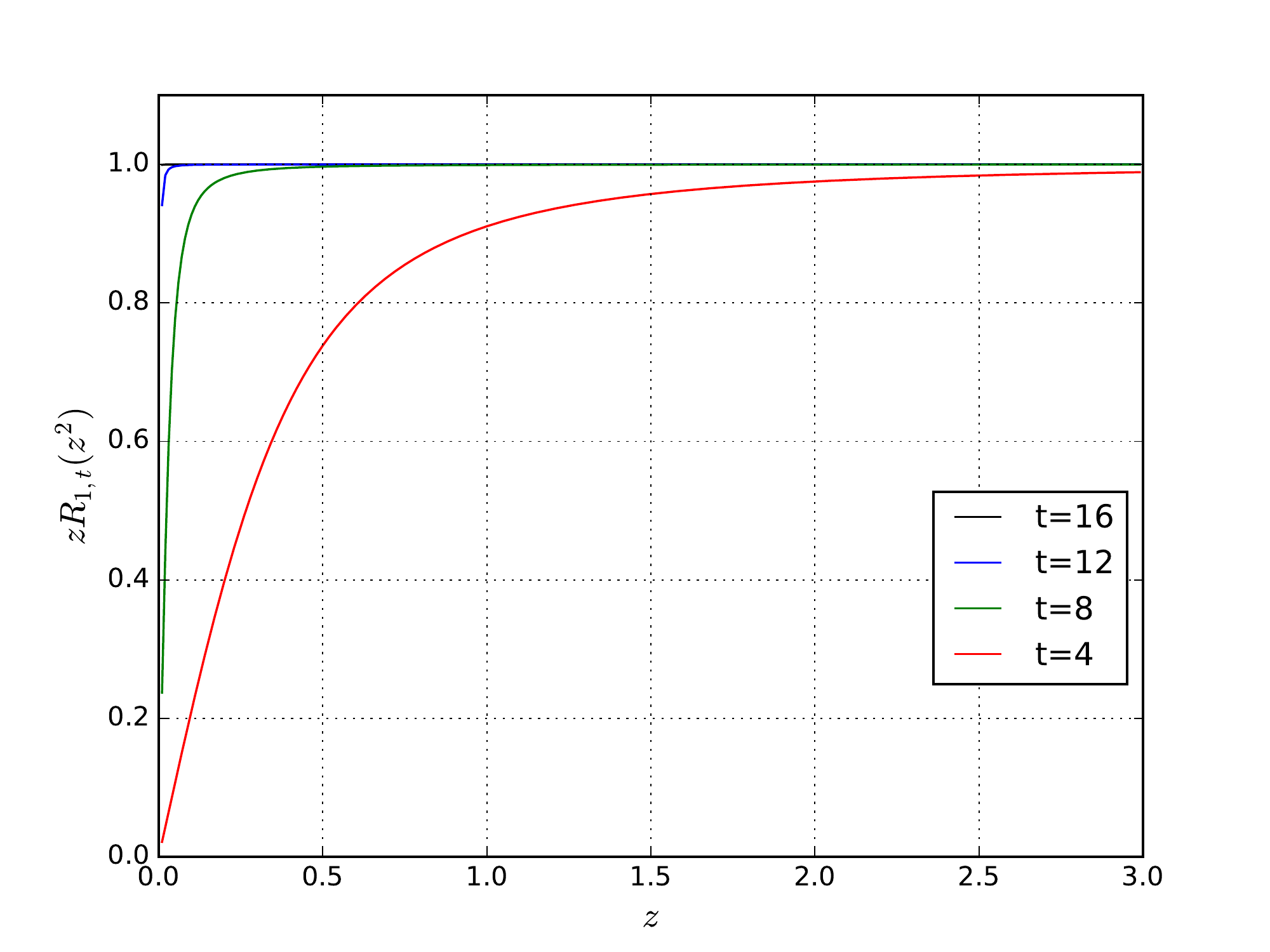}
  \caption{The function $zR_{0,t}(z^2) \simeq 1$ for various values of
    the truncation order $t.$ The truncated products $R_{0,t}$ derive
    from a Zolotarev rational polynomial $R(z^2) \simeq 1/\sqrt{z^2}$ of
    order $n=20$ over the range $z \in [5\times 10^{-5},3].$ The graph
    demonstrates that truncations of the ordered product also
    approximate the inverse square root.}
  \label{fig:tzolo}
\end{figure}%
This property is demonstrated in Figure~\ref{fig:tzolo}, where we plot
the functions $zR_{0,t}(z^2) \simeq 1$ for $z > 0,$ where $R_{0,t}$
are the truncations of the $n=20$ Zolotarev approximation specified in
Table~\ref{tab:zolo}. We see that for all choices of $t$ the
approximation is good at large values of $z,$ while the range over
which the approximation is valid gets successively better as the
truncation order $t$ increases. We observe at $t=16$ the approximation
error is negligible over almost the entire range of $z.$ Hence we have
shown that truncations of the ordered product successfully capture the
high frequency dynamics.
Next we consider the correction term which is simply the remaining
factors associated with a given truncation order $t,$
\be R_{t,n}(K) = \prod_{k=t+1}^n \frac{(K+a_k)}{(K+b_k)} \simeq 1. \ee
\begin{figure}[tb] 
  \centering
  \includegraphics[width=0.7\textwidth]{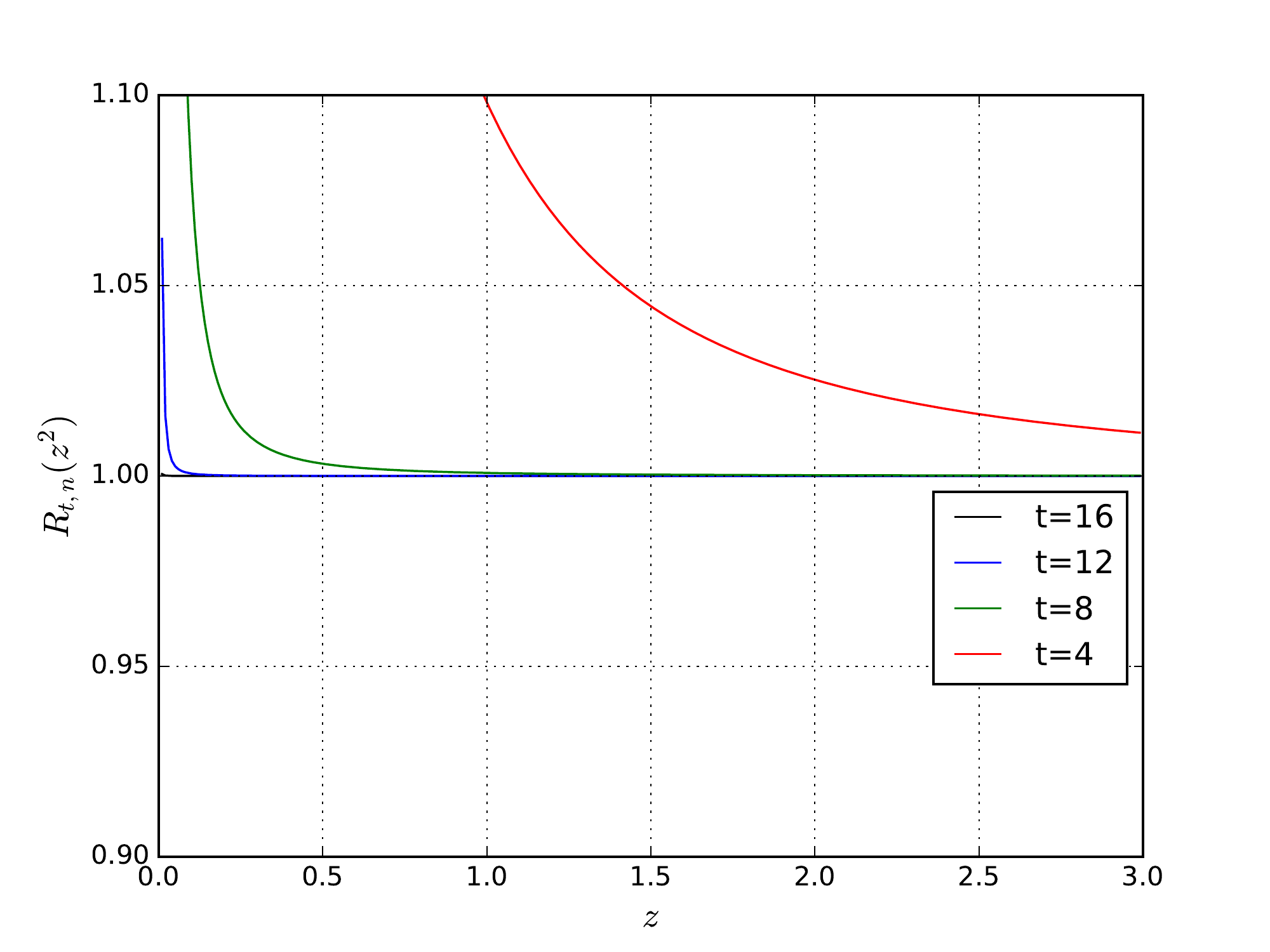}
  \caption{The correction term $R_{t,n}(z^2) \simeq 1$ for various values of
    the truncation order $t.$ The remainders $R_{t,n}$ derive
    from a Zolotarev rational polynomial $R(z^2) \simeq 1/\sqrt{z^2}$ of
    order $n=20$ over the range $z \in [5\times 10^{-5},3].$ The graph
    demonstrates that truncating the ordered product yields remainders
    that approximate unity.}
  \label{fig:tcorr}
\end{figure}%
Note that, due to the ordering of the product, for values of $k >> 1$
we have that $a_k \simeq b_k$ are both small, and as the order of the
polynomials in the numerator and denominator are equal, the correction
term is an approximation to unity. This is very clearly demonstrated
in Figure~\ref{fig:tcorr}, which shows the correction terms
$R_{t,n}(z^2)$ arising from different choices of the truncation order
$t$ for the Zolotarev approximation from Table~\ref{tab:zolo}. The
bounds on the vertical axis are set at a 10\% deviation from 1. Even
at $t = 8,$ the correction term is approximately one over a large range,
while at $t=16$ the remainder is indistinguishable from unity at the
displayed scale. This is significant, as it implies that the
correction term will have a correspondingly small molecular dynamics
force and thus can be placed on a coarse integration scale.

Based on the above observations, we define the \emph{truncated ordered
product (top)} filtering scheme for RHMC, denoted by tRHMC, such that
\be S_{tRHMC} = \phi_1^\dag R_{0,t}(K) \phi_1 + \phi_2^\dag R_{t,n}(K) \phi_2 \ee
is the tRHMC pseudofermion action. Here we point out a significant
advantage of the tRHMC scheme is that, as both terms are simply
rational polynomials, it is very easy to implement using preexisting
RHMC code.

It is also worth noting that partitioning the ordered product form
of the rational polynomial is distinct from partitioning the sum over
poles, in which different poles in the expansion of $R(K)$ are simply
placed on different time scales. The reason is that the sum over poles
simply distributes the inversions based on their expense, and does not
share the beneficial properties of the tRHMC scheme mentioned above,
namely that by also approximating $K^{-\frac{1}{2}},$ $R_{0,t}$ acts
as a high pass filter. A further computational efficiency for the
tRHMC scheme may be realised on some hardware architectures due to the
memory bandwidth benefits of evaluating multiple small order rational
polynomial terms, rather than a single term of full order.


As has been done in the two flavour case, hierarchical filtering
schemes can be defined for the single flavour filters that we have
proposed. In the case of PF-RHMC, we can add a higher order polynomial
$Q(K), (q > p)$ which approximates $P(K)^{-1}K^{-1/2}$ as a secondary
filter, which results in the 2PF-RHMC action,
\be S_{2PF-RHMC} = \phi_1^\dag P(K) \phi_1 + \phi_2^\dag Q(K) \phi_2 + \phi_3^\dag R(K)[P(K)Q(K)]^{-1} \phi_3. \ee
For the tRHMC scheme, we can add a second truncation point in the order product, such that the 2tRHMC action is given by
\be S_{2tRHMC} = \phi_1^\dag R_{0,t_1}(K) \phi_1 + \phi_2^\dag R_{t_1,t_2}(K) \phi_2 + \phi_3^\dag R_{t_2,n}(K) \phi_3. \ee
%

\section{Characteristic scale tuning}

Using a fully generalised integration scheme~\cite{Kamleh:2011dc,Haar:2016bwe}, it is possible
to place each term in the molecular dynamics action on an independent
scale. Typically, the step sizes $h_i$ associated with a given action
term $S_i$ can be tuned by force balancing, where given some measure
$F_i$ of the molecular dynamics forces, for each term we attempt to
balance the product of the step size and the force such that $F_i h_i \sim \mathrm{constant}.$
%
%
Here, $F_i$ can be the average or maximum force associated with a
given term. The maximum tends to prove a better choice, as it is more
closely linked with the force variance.

However, in the case of tRHMC, for higher order truncations, the force
due to the remainder term is very small. This makes the force-balanced
approach to tuning the step sizes unfavourable. Noting that for
filtered actions the acceptance rate is primarily determined by the
coarsest scale, we adopt an alternative approach to tuning using the
characteristic scale. The coarsest scale $h_0$ is tuned by fitting the
acceptance probability with the complementary error function, $\rho_{\rm acc} \approx \erfc(h_0^2 c^2),$
%
%
where the characteristic scale $c$ is the only fit
parameter. With the coarsest scale $h_0$ excluded, the remaining
scales $h_1 > h_2 > h_3 \ldots$ are set using the force balancing
relation 
to achieve the target acceptance rate.

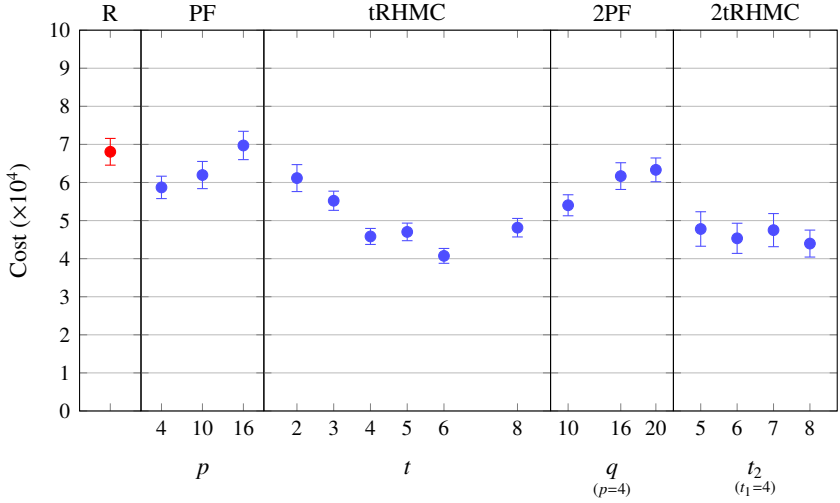
\begin{figure}[tb] 
  \centering
\newcommand{\widthdelta}{10cm/(1.5+3+7+3+4)}

\begin{tikzpicture}[baseline, trim axis group left]

\begin{groupplot}[
	group style={
		group size=6 by 1,
		horizontal sep=0pt,
		yticklabels at=edge left,
		},
	ymin=0, ymax=100000,
	footnotesize,
	height=5cm,
	scaled y ticks=base 10:-4,
	ytick scale label code/.code={},
	y errors,
	ymajorgrids,
	scale only axis=true,
	xlabel style={
		at={(0.5,-0.15)},
		anchor=mid,
	},
	normal plot,
	]


\nextgroupplot[
	title={R},
	ylabel={Cost ($\times 10^4$)},
	xtick=data,
	xticklabel=\empty,
	width=1.5*\widthdelta,
	only marks,
]

\renewcommand{\matrixopfile}{\datadir/witers_j1350.txt}
	
\addplot+[mark1]
	table[
		x=ID,
		y expr={\thisrow{iter_SF} + \thisrow{iter_F1} + \thisrow{iter_F2}},
		y error expr={\thisrow{iter_tot_err}*\autocorr},
	]
	{\matrixopfile};


\nextgroupplot[
	title={PF},
	xlabel={$p$},
	only marks,
	xtick=data,
	width=3*\widthdelta,
	enlarge x limits=0.25,
]

\renewcommand{\matrixopfile}{\datadir/witers_j135X_cs.txt}
	
\addplot+[mark2]
	table[
		x=p,
		y expr={\thisrow{iter_SF} + \thisrow{iter_F1} + \thisrow{iter_F2}},
		y error expr={\thisrow{iter_tot_err}*\autocorr},
	]
	{\matrixopfile};


\nextgroupplot[
	title={tRHMC},
	xlabel={$t$},
	only marks,
	xtick=data,
	width=7*\widthdelta,
	enlarge x limits=0.15,
]

\renewcommand{\matrixopfile}{\datadir/witers_j138X_cs.txt}

\addplot+[mark2]
	table[
		x=t,
		y expr={\thisrow{iter_SF} + \thisrow{iter_F1} + \thisrow{iter_F2}},
		y error expr={\thisrow{iter_tot_err}*\autocorr},
	]
	{\matrixopfile};


\nextgroupplot[
	title={2PF},
	width=3*\widthdelta,
	xlabel={$\underset{(p=4)}{q}$},
	xtick=data,
	only marks,
	enlarge x limits=0.2,
]

\renewcommand{\matrixopfile}{\datadir/witers_j136X_cs.txt}
	
\addplot+[mark2]
	table[
		x=q,
		y=iter_tot,
		y error expr={\thisrow{iter_tot_err}*\autocorr},
	]
	{\matrixopfile};


\nextgroupplot[
	title={2tRHMC},
	xlabel={$\underset{(t_1=4)}{t_2}$},
	width=4*\widthdelta,
	only marks,
	xtick=data,
	enlarge x limits=0.25,
]

\renewcommand{\matrixopfile}{\datadir/witers_j1380X_t4_cs.txt}

\addplot+[mark2]
	table[
		x=t2,
		y=iter_tot,
		y error expr={\thisrow{iter_tot_err}*\autocorr},
	]
	{\matrixopfile};







\end{groupplot}
\end{tikzpicture}
  \caption{The cost function $C = N_{\rm mat}/\rho_{\rm acc}$ for the different algorithms used in this study. The red point is for vanilla RHMC, while the blue points are for the filtering schemes using characteristic scale tuning. From left to right, first we have the single filter results at various values of their respective order parameters, namely PF-RHMC for different polynomial order $p,$ and tRHMC for different truncation order $t.$ Next we have the two filter results, with 2PF-RHMC at fixed $p=4$ and various $q$ for the polynomial orders, followed by 2tRHMC at fixed $t_1=4$ and various $t_2$ for the truncation orders.}
  \label{fig:summary}
\end{figure}

\section{Results}

All runs were performed using the open source BQCD software
package~\cite{Nakamura:2010qh,Haar:2017ubh}. Both the PF-RHMC and tRHMC filtering methods are
built into the latest release, which is available at the following
link:
\begin{center}
\href{https://www.rrz.uni-hamburg.de/services/hpc/bqcd}{www.rrz.uni-hamburg.de/services/hpc/bqcd}
\end{center}
To compare the different filtering algorithms proposed herein, we
start with a thermalised $16^3 \times 32$ lattice with $N_f = 2$
flavours of Wilson fermions from a previous study~\cite{Haar:2016bwe}. The Wilson
gauge action is used with $\beta = 5.6,$ with hopping parameter
$\kappa = 0.15825,$ providing a pion mass of $m_\pi \sim
400~\mathrm{MeV}$ and a lattice spacing of $a \sim 0.08~\mathrm{fm}.$

The various RHMC simulations are then performed with (degenerate)
$1+1$ single-flavour pseudofermions. As both flavours are treated
equally, they have degenerate force and fermion matrix multiplication
count distributions. We tune the step-sizes using the characteristic
scale method above such that the acceptance rate $\rho_{acc} \sim
0.75$. The rational polynomial used is the Zolotarev approximation
with order $n=20$ and range $[5 \times 10^{-5}, 3]$ given in
Table~\ref{tab:zolo}. The polynomial filters used in the PF-RHMC runs are the
Chebyshev approximations to $K^{-1/2}$ (or $P(K)^{-1}K^{-1/2}$ in the
two filter case) with range $[5 \times 10^{-5}, 3]$. The cost function
we use to compare our methods is the same as in~\cite{Haar:2016bwe},
namely
\be C = N_{\rm mat}/\rho_{\rm acc}, \ee
where the matrix multiplication count $N_{\rm mat}$ is the total for both flavours.

Figure~\ref{fig:summary} compares the cost function for the various
filtering methods described herein. Polynomial-filtered RHMC provides
a small benefit over vanilla RHMC. It is clear that tRHMC provides the
best performance improvement, at a speedup of around 1.75. The optimal
point for tRHMC and 2tRHMC is similar, but the two truncation form is
less sensitive to the choice of the truncation parameter.

In summary, we have seen that the tRHMC filtering scheme coupled with
characteristic scale tuning works well cross a wide range of
truncation orders. The tRHMC algorithm is provided by BQCD, but is
also very easy to implement in existing RHMC codes. Multiple
truncation filters can be used, and this should prove to be useful at
light quark masses. The full details of our study will be published
elsewhere.

%


\end{document}